\documentclass{cimento}

\usepackage{graphicx}  
\usepackage{graphbox}

\title{Continuum emission enhancements and penumbral changes \ETC \\ observed during flares by \textit{IRIS}, ROSA, and \textit{Hinode}}
\author{F.~Zuccarello\from{ins:a}\ETC,
S.L.~Guglielmino\from{ins:a},
V.~Capparelli\from{ins:a},
M.~Mathioudakis\from{ins:b},
P.~Keys\from{ins:b},
L.~Fletcher\from{ins:c},
S.~Criscuoli\from{ins:d},
M.~Falco\from{ins:e}
        \atque
M.~Murabito\from{ins:f}}
\instlist{\inst{ins:a} Dipartimento di Fisica e Astronomia ``Ettore Majorana'' -- Sez.~Astrofisica, Universit\`{a} degli Studi di Catania, Via S.~Sofia 78, I-95123 Catania, Italy
	\inst{ins:b} Astrophysics Research Centre, School of Mathematics \& Physics, Queen's University Belfast, Belfast, BT7 1NN, UK
	\inst{ins:c} SUPA, School of Physics \& Astronomy, University of Glasgow, G12 8QQ, Scotland, UK
	\inst{ins:d} NSO -- National Solar Observatory, Sacramento Peak - Box 62, Sunspot NM 88349, USA
	\inst{ins:e} INAF -- Oss.~Astrofisico di Catania, Via S.~Sofia 78, I-95123 Catania, Italy
	\inst{ins:f} INAF -- Oss.~Astronomico di Roma, via Frascati 33, I-00078 Monte Porzio Catone, Italy}

\begin{document}

\maketitle

\begin{abstract}
In this paper we describe observations acquired by satellite instruments (\textit{Hinode}/SOT and \textit{IRIS}) and ground-based telescopes (ROSA@DST) during two consecutive C7.0 and X1.6 flares occurred in active region NOAA 12205 on 2014 November 7. The analysis of these data show the presence of continuum enhancements during the evolution of the events, observed both in ROSA images and in \textit{IRIS} spectra. Moreover, we analyze the role played by the evolution of the $\delta$ sunspots of the active region in the flare triggering, indicating the disappearance of a large portion of penumbra around these sunspots.
\end{abstract}

\section{Introduction}
During solar flares, electromagnetic radiation from radio waves to $\gamma$ rays can be emitted due to the conversion of magnetic energy as a consequence of magnetic reconnection (\cite{Zuccarello:13}, \cite{Romano:15}, \cite{Guglielmino:16}). Enhancements in the continuum at visible wavelengths (white-light [WL] flares, see, e.g., \cite{neidig83}), as well as in the FUV/NUV passbands may be observed, especially in the case of the most energetic flares. Moreover, the strong energy release occurring during reconnection can give rise to a rearrangement of the magnetic field at the photospheric level, leading in some cases to morphological changes in the penumbrae of sunspots \cite{Song:16}. In this paper we show some examples of these processes occurring during two consecutive flares.

\begin{figure}[t]
	\centering
	\hspace*{-3ex}
	\includegraphics[align=c,trim=20 100 20 120, clip, scale=0.23]{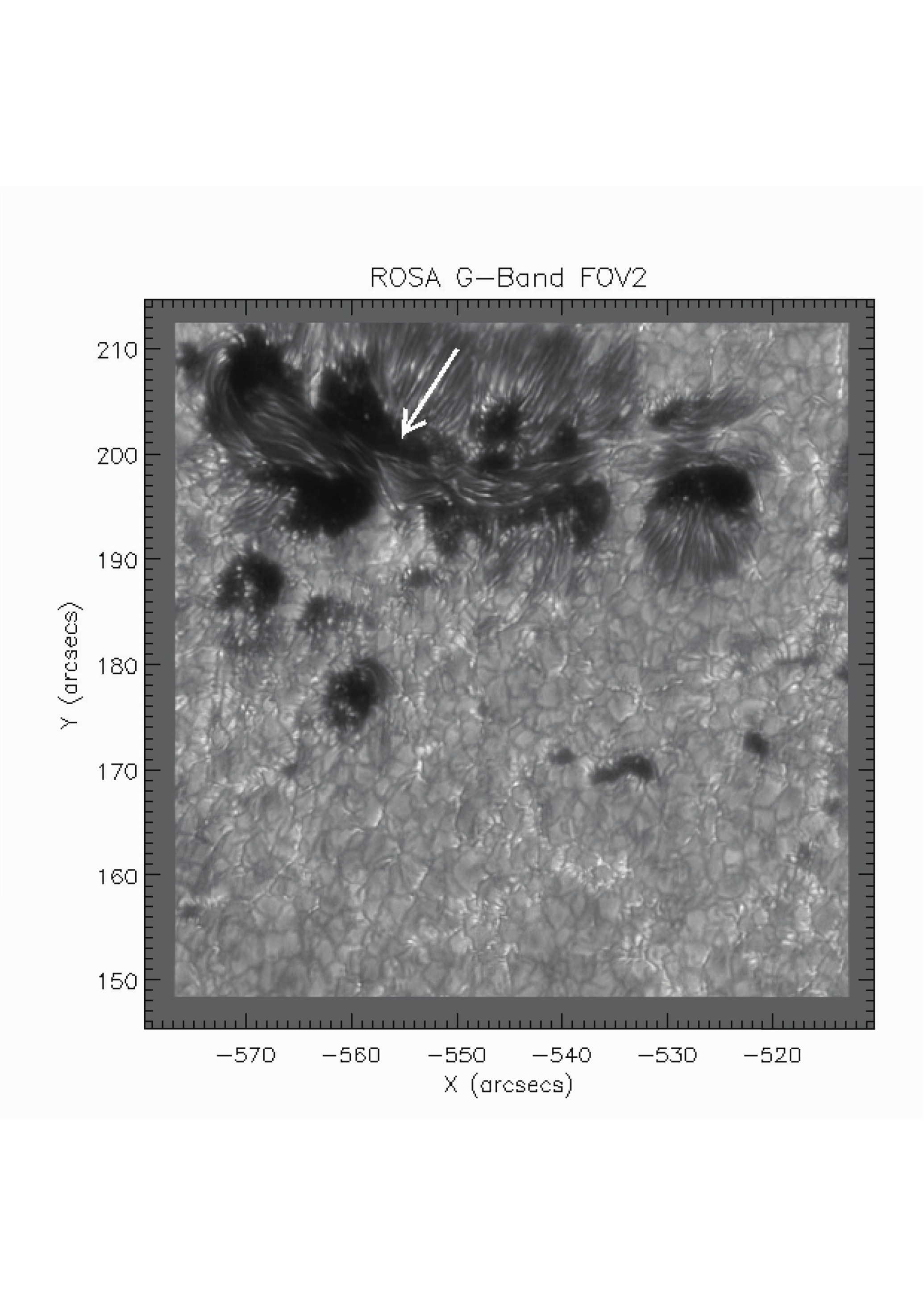}%
	\hspace*{-6ex}	
	\includegraphics[align=c,trim=20 0 30 30, clip,scale=0.23]{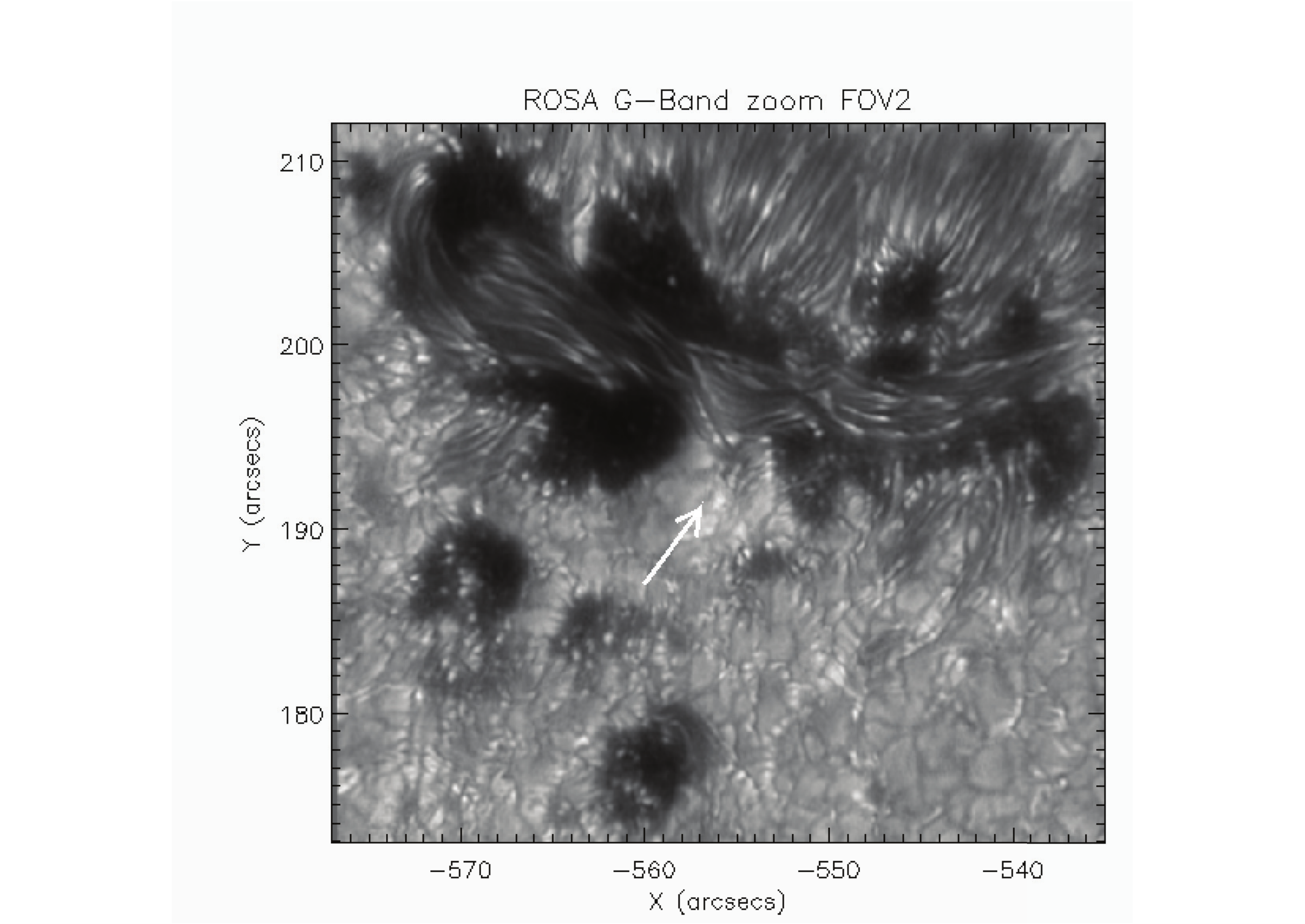}%
	\hspace*{-4ex}
	\includegraphics[align=c, trim=40 20 1 1, clip, scale=0.565]{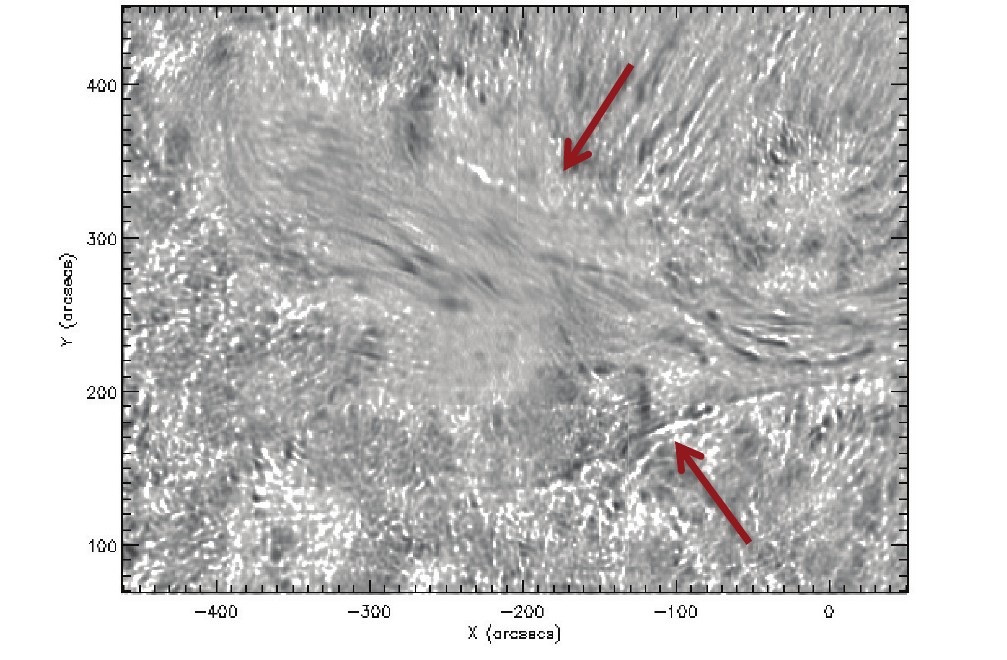}
	\caption{\textit{Left}: ROSA \textit{G}-band image showing the entire FOV2, during the occurrence of the X1.6 flare. The white arrow indicates the southern $\delta$ spot of the AR. \textit{Middle}: Zoomed image showing the location of a ribbon observed in the \textit{G}-band (white arrow) at 17:22~UT (i.e., few minutes before the X1.6 flare peak). \textit{Right}: \textit{G}-band difference image showing part of FOV2, with the location of the WL ribbons (red arrows). North is on the top, West is to the right. \label{fig:ROSA}}
\end{figure}

\section{Observations}
On 2014 November~7, active region (AR) NOAA~12205 was observed during a coordinated observing campaign carried out using the ground-based ROSA (Rapid Oscillations in the Solar Atmosphere, \cite{jess10}) instrument at the Dunn Solar Telescope (DST) and the \textit{IRIS} (Interface Region Imaging Spectrograph, \cite{depontieu}) satellite. ROSA@DST acquired data simultaneously in the Blue continuum at 4170~\AA{} and in the \textit{G} band at 4305.5~\AA{}, with a diffraction limited spatial sampling of $0.069^{\prime\prime}$/pixel in two different fields of view (both $69^{\prime\prime} \times 69^{\prime\prime}$). \textit{IRIS} acquired SJI images from 16:07 to 16:57~UT at 3 wavelengths (C II at 1330~\AA, Mg II~k at 2796~\AA, Mg II wing at 2832~\AA) and used the large 4-step coarse raster mode. High-resolution data acquired in the \textit{G} band between 15:15~UT and 19:06~UT by the Solar Optical Telescope (\cite{Tsuneta:08}) aboard the \textit{Hinode} satellite (\cite{Kosugi:07}) were also used. 

\begin{figure}[t]
	\centering
	\includegraphics[trim=20 40 0 20, clip, scale=0.373]{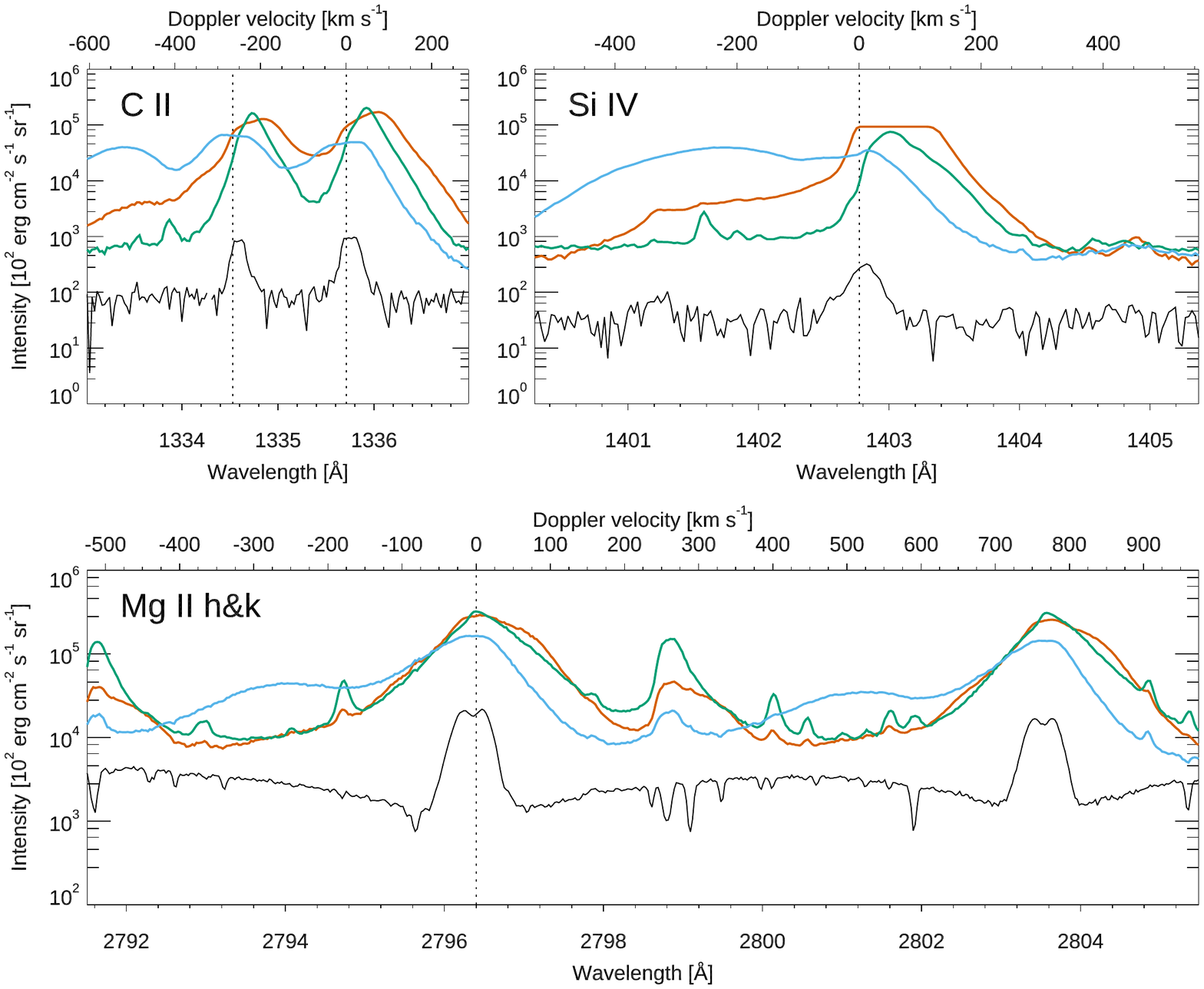}
	\hspace*{-12ex}
	\includegraphics[trim=545 50 120 255, clip, scale=0.595]{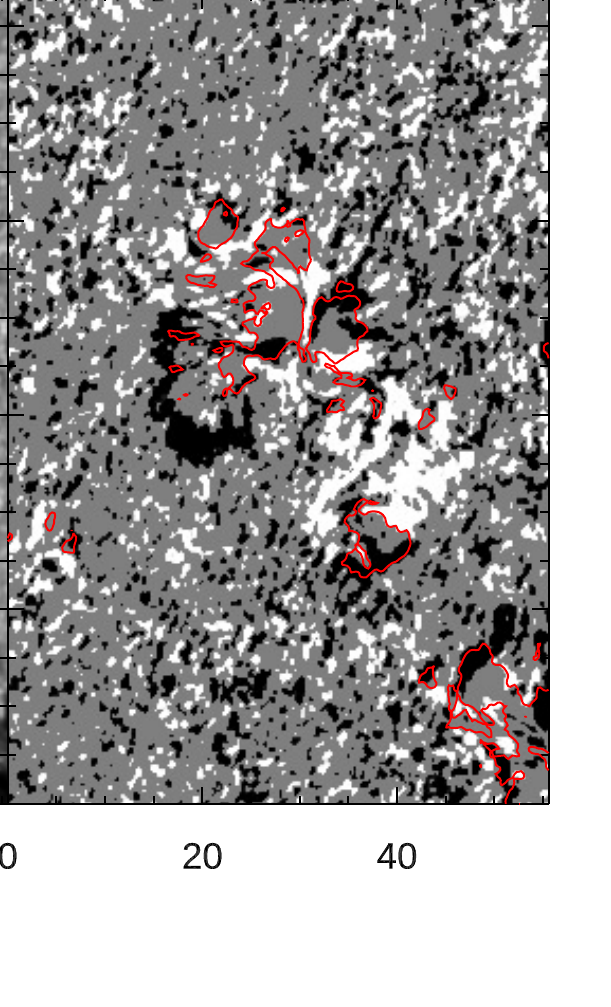}
	\caption{\textit{Left}: Plots of the average intensity as a function of wavelength in three of the \textit{IRIS} spectral windows in the pixels at raster positions (3,[537:539]; blue), (3,[543:545]; orange), and (3,[558:560]; green) at 16:56:47~UT. For comparison, the black line is the average intensity calculated at the same time in a quiet-Sun region. \textit{Right}: Difference image between \textit{G}-band filtergrams acquired by \textit{Hinode}/SOT at 18:36:40 UT and 15:15:04 UT. White (black) areas indicate regions with penumbral enhancement (decay). Red contours indicate the umbral boundary at 15:15:04 UT.
	\label{fig:iris_continua}}
\end{figure}

\section{Results}
AR NOAA 12205 was characterized by the presence of two $\delta$ spots: one, located on the northern side of the AR, comprised in the ROSA FOV1, showed the presence of penumbral filaments running almost parallel to the umbrae edges; the other, included in the ROSA FOV2 and located in the southern part of the AR, was characterized by a magnetic inversion line almost parallel to the equator (see Figure~\ref{fig:ROSA}, \textit{left}).
Two flares took place in the AR during the observations: a C7.0 flare (peak at 16:39~UT) and an X1.6 flare, starting at 16:53~UT, peaking at 17:26~UT, ending at 18:34~UT. Few minutes before the peak of the X1.6 flare (17:22~UT), the analysis of ROSA images relevant to FOV2 indicates that both in the \textit{G} band and in the 4170~\AA{} continuum, it is possible to detect WL ribbons close to the southern $\delta$ spot (see, e.g., Figure~\ref{fig:ROSA}, \textit{middle}), separating with a velocity of $\sim10 \,\mathrm{km\,s}^{-1}$. In particular, the southernmost ribbon is clearly visible in the images and movies obtained in these wavelength ranges, while the northern ribbon can be distinguished only using the difference images (see, e.g., Figure~\ref{fig:ROSA}, \textit{right}). 

In Figure~\ref{fig:iris_continua} (\textit{left}) we display the radiometric calibrated intensities for selected pixels along the \textit{IRIS} slit at 16:56:47 UT. The blue, orange and green curves show the intensity in correspondence of a ribbon observed in the northern $\delta$ spot. We can see that in all the channels there is an intensity enhancement also in the continuum region. Moreover, the blue pixel exhibits a very prominent bump in the blue wing of the Si IV 1402~\AA{} line, but also in the blue wings of the C II 1334 and 1336, Mg II~h and~k lines.

Using filtergrams obtained by \textit{Hinode}/SOT in the \textit{G} band, we obtained the difference image displayed in Figure~\ref{fig:iris_continua} (\textit{right}), showing areas of permanent penumbral decay (black) and enhancement (white) during the evolution of the observed flares. 

\acknowledgments
The authors acknowledge support by the Universit\`a degli Studi di Catania (Piano per la Ricerca 2016-2018 -- Linee di intervento~1-2) and by the European Union's H2020 programme under grant agreement no.~739500 (PRE-EST project).

\end{document}